\documentclass{article}
\usepackage{fullpage,amsthm,amsfonts,amsmath}
\usepackage{amsmath}
\usepackage{amssymb}
\usepackage{algorithm}
\usepackage{algorithmic}
\usepackage{verbatim}

\newtheorem{theorem}{Theorem}
\newtheorem{lemma}{Lemma}
\newtheorem{corollary}{Corollary}
\newtheorem{proposition}{Proposition}

\newcommand{\ket}[1]{\lvert #1 \rangle}

\newcommand{\size}[1]{\lvert #1 \rvert}

\newcommand{\C}{\mathbb{C}}
\DeclareMathOperator{\supp}{\mathsf{supp}}

\newcommand{\suppress}[1]{}
\newcommand{\journal}[1]{}

\def\Z{{\mathbb Z}}

\newcommand{\HSP}{\mbox{\textsc{HSP}}}
\newtheorem{fact}{Fact}
\newtheorem{claim}{Claim}

\begin{document}

\title{An efficient quantum algorithm for  the hidden subgroup problem
in nil-$2$ groups
\thanks{
Research supported by the European Commission IST Integrated Project
Qubit Applications (QAP) 015848, the OTKA grants T42559 and T46234,
the NWO visitor's grant Algebraic Aspects of Quantum Computing,
and by the
ANR Blanc AlgoQP grant of the French Research Ministry.
}}

\author{
G\'abor Ivanyos\footnotemark[2]
\and
Luc Sanselme\footnotemark[3]
\and
Miklos Santha\footnotemark[4]
}

\maketitle

\renewcommand{\thefootnote}{\fnsymbol{footnote}}

\footnotetext[2]{SZTAKI, Hungarian Academy of Sciences,
H-1111 Budapest, Hungary.
{\tt ivanyos}{\tt @sztaki.hu}}

\footnotetext[3]{UMR 8623 Universit\'e Paris--Sud
91405 Orsay, France.
{\tt sanselme@lri.fr}
}

\footnotetext[4]{CNRS--LRI, UMR 8623 Universit\'e Paris--Sud
91405 Orsay, France.
{\tt santha@lri.fr}
}

\begin{abstract}
In this paper we extend the algorithm  for extraspecial groups in~\cite{iss07}, and
show that the hidden subgroup problem in nil-$2$ groups,
that is in groups of nilpotency class at most 2, can be solved efficiently by a 
quantum procedure. The algorithm presented here has several additional features.
It contains a powerful classical reduction for the hidden subgroup problem in
nilpotent groups of constant nilpotency class to the specific case where the group is
a $p$-group of exponent $p$ and the subgroup is either trivial or cyclic. This reduction might
also be useful
for dealing with groups of higher  nilpotency class. The quantum part of the algorithm
uses well chosen group actions based on some automorphisms of nil-$2$ groups. The right
choice of the actions requires the solution of a system of quadratic and linear equations.
The existence of a solution is guaranteed by the Chevalley-Warning theorem, 
and we prove
that it can also be found efficiently.

\suppress{
Extraspecial groups form a remarkable subclass of $p$-groups.
They are also present in quantum information theory, in particular
in quantum error correction. We give here a polynomial time
quantum algorithm for finding hidden subgroups in extraspecial groups.
Our approach is quite different from the recent
algorithms presented in \cite{rrs05} and \cite{bcv05}
for the Heisenberg group, the extraspecial $p$-group of size
$p^3$ and exponent $p$. Exploiting certain nice automorphisms of the
extraspecial groups we define specific
group actions which are used to reduce the problem to hidden subgroup instances
in abelian groups that can be dealt with directly.
}
\end{abstract}

\newpage

\section{Introduction}
Efficient solutions to some cases of the hidden subgroup problem (HSP),
a paradigmatic group theoretical problem, constitute probably the most notable
success of quantum computing. The problem consists in finding a subgroup $H$ in a
finite group $G$ hidden by some function which is constant on each coset of $H$ and is 
distinct in different cosets. The hiding function can be accessed by an oracle, 
and in the overall complexity of an algorithm, a query counts as a single computational step.
To be efficient, an algorithm has to be polylogarithmic in the order of $G$. While classically
not even query efficient algorithms are known for the HSP, it can be solved efficiently in abelian
groups by a quantum algorithm. A detailed description of the so called standard algorithm
can be found for example in~\cite{mos99}. The main quantum tool of this algorithm is 
Fourier sampling, based on the efficiently implementable  Fourier transform in abelian groups.
Factorization and discrete  logarithm~\cite{sho97}
are special cases of  this solution.
 
After the settling of the abelian case, substantial research was devoted to the HSP
in some finite non-abelian groups. Beside being the natural generalization of the abelian case,
the interest of  this problem is enhanced by the fact, that important algorithmic problems, such as
graph isomorphism, can be cast in this framework.
The standard
algorithm has been extended to some non-abelian groups
by R\"otteler and Beth~\cite{rb98}, Hallgren, Russell and Ta-Shma~\cite{hrt03},
Grigni, Schulman, Vazirani and Vazirani~\cite{gsvv01} and
Moore, Rockmore, Russell and Schulman~\cite{mrrs04}. For the Heisenberg group,
Bacon, Childs and van~Dam~\cite{bcv05} used the pretty good
measurement to reduce the HSP to some matrix sum problem that they could solve
classically. Ivanyos, Magniez and Santha~\cite{ims03} and
Friedl, Ivanyos, Magniez, Santha and Sen~\cite{fimss03} have efficiently reduced
the  \HSP{} in some non-abelian groups to \HSP{} instances in abelian groups
using classical and quantum group theoretical tools, but not the
non-abelian Fourier transform. This latter approach was used recently by Ivanyos, Sanselme and
Santha~\cite{iss07} for extraspecial groups.

In this work we extend the class of  groups where the HSP is efficiently solvable by a
quantum algorithm to nilpotent groups of nilpotency class at most 2 (shortly nil-$2$ groups).
These are groups whose lower (and upper) central series are of length at most 2. Equivalently,
a  group is nil-$2$ group if  the derived group is a subgroup of the  center. Nilpotent groups form
a rich subclass of solvable groups, they contain for example all (finite) $p$-groups. Extraspecial 
groups are, in particular, in nil-$2$ groups.
Our main result is:
\begin{theorem}\label{theorem:main}
Let $G$ be a nil-$2$ group, and
let us given an oracle $f$ which
hides the subgroup $H$ of $G$. Then there is an efficient quantum procedure which
finds $H$.
\end{theorem}

The overall structure of the algorithm presented here is closely related to 
the algorithm in~\cite{iss07} for 
extraspecial groups, but has also several additional features. The quantum part of the
algorithm is restricted to specific nil-$2$ groups, which are also $p$-groups and are of
exponent $p$. It consists essentially in the creation of a quantum hiding procedure
(a natural quantum generalization of a hiding function) for the subgroup $HG'$ of $G$.
The procedure uses certain automorphisms of the groups to define some appropriate group actions,
and is analogous to what have been done in~\cite{iss07}  for extraspecial $p$-groups
of exponent $p$.

While dealing with extraspecial $p$-groups of exponent $p$ basically
solves the HSP for all extraspecial groups (the case of remaining groups, of exponent $p^2$,
easily reduces to groups of exponent $p$), this is far from being true for nil-$2$ groups.
Indeed, one of the main new features of the current algorithm is a classical reduction
of the HSP in nil-$2$ groups to the HSP in nil-$2$ $p$-groups of exponent $p$, 
where moreover the 
hidden subgroup is either trivial or of cardinality $p$. In fact, our result is much more general: 
we prove an analogous reduction in nil-$k$ groups for any constant $k$. We believe that this
general reduction might be useful for designing efficient quantum algorithms for the HSP
in groups of higher nilpotency class.

Our second main novel feature concerns the quantum hiding procedure. While in extraspecial
groups it was reduced to the efficient solvability of a single quadratic and a 
single linear equation modulo $p$, here we look for a nontrivial solution of a homogeneous
system of  $d$  quadratic and $d$ linear equations, where $d$ can be any
integer. The reason for this is that while in extraspecial groups the derived subgroup is one dimensional,
in nil-$2$ groups we have no a priori bound on its dimension.
If the number of 
variables is superior to the global degree of the 
system then
the solvability itself is an
immediate consequence of the Chevalley-Warning theorem ~\cite{c36, w36}. In fact, we are
in presence of a typical example of Papdimitriou's complexity class of total functions~\cite{mp91}: 
the number of solutions is divisible by $p$ and therefore there is always a nontrivial one.
Our result is that if the number of variables 
is sufficiently large, more precisely is of $O(d^3)$, then we can 
also find a nontrivial solution in polynomial time.

The structure of the paper is the following.
In Section~\ref{prelim} we shortly describe the extension of the standard algorithm for quantum hiding
procedures, and then we discuss some basic properties of nilpotent groups, in particular nil-$2$
$p$-groups of exponent $p$. Section~\ref{section:redclas} contains the description of the classical
reduction of the HSP in groups of constant nilpotency class to instances where the group is
also $p$-group of exponent $p$, and the subgroup is either trivial or cyclic of order $p$
(Theorem~\ref{theorem:redclas}). Section~\ref{section:algorithm} gives the description of the quantum 
algorithm in nil-$2$ $p$-groups of exponent $p$: Theorem~\ref{theorem:threelemmas}
briefly describes the reduction to the design of an efficient hiding procedure for $HG'$, and 
Theorem~\ref{theorem:HG'} proves the existence of such a procedure. 
Finally Section~\ref{section:equation} gives the proof of Theorem~\ref{theorem:equation},
the efficient solvability of the system of quadratic and linear equations.
The proof of Theorem~\ref{theorem:main} follows from Corollary~\ref{corollary:redclas} and
Theorems~\ref{theorem:threelemmas} and~\ref{theorem:HG'}.

\section{Preliminaries}\label{prelim}

\subsection{Extension of the standard algorithm for the abelian \HSP}

We will use standard notions of quantum computing for which one can consult for
example~\cite{nc00}.
For a finite set $X$, we denote by $\ket{X}$ the uniform superposition
$\frac{1}{\sqrt{|X|}}\sum_{x \in X} \ket{x}$ over $X$.
For a superposition $\ket{\Psi}$, we denote by $\supp(\ket{\Psi})$ the support of $\ket{\Psi}$,
that is the set of basis
elements with non-zero amplitude.

The standard algorithm for the abelian $\HSP$ repeats
polynomially many times the Fourier sampling
involving the same hiding function, to obtain 
in each iteration a random element from the subgroup orthogonal to the hidden subgroup.
In fact, for the repeated Fourier samplings, the existence of a common hiding function can be
relaxed in several ways. Firstly, in different iterations different hiding functions can be used, 
and secondly, classical hiding functions can be replaced by quantum hiding functions.
This was formalized in \cite{iss07}, and we recall here the precise definition.

A set of vectors $\{ \ket{\Psi_g} : g \in G \}$ from some Hilbert space $\cal{H}$ is
a {\em hiding set} for
the subgroup $H$ of $G$ if
\begin{itemize}
\item
$\ket{\Psi_g}$ is a unit vector for every $g \in G$,
\item
if $g$ and $g'$ are in the same left coset of $H$ then $\ket{\Psi_g} = \ket{\Psi_{g'}}$,
\item
if $g$ and $g'$ are in different left cosets of $H$ then $\ket{\Psi_g}$ and $\ket{\Psi_{g'}}$
are orthogonal.
\end{itemize}
A quantum procedure is {\em hiding} the subgroup $H$ of $G$ if
for every $g_1, \ldots, g_N \in G$, on input
$\ket{g_1} \ldots \ket{g_N}\ket{0}$ it outputs
$\ket{g_1} \ldots \ket{g_N}\ket{\Psi_{g_1}^1} \ldots \ket{\Psi_{g_N}^N}$, where
$\{ \ket{\Psi_g^i} : g \in G \}$ is a hiding set for $H$ for all $1 \leq i \leq N$.

The following fact whose proof is immediate from Lemma 1 in \cite{ims03} recasts
the existence of the standard algorithm for the abelian $\HSP$ in the context of hiding sets. 
\begin{fact}\label{fact:hsp}
Let $G$ be a finite abelian group. If there exists an efficient quantum procedure
which hides the subgroup $H$ of $G$ then there is an efficient quantum algorithm for
finding $H$.
\end{fact}

\subsection{Nilpotent groups}
Let $G$ be a finite group. For two elements $g_1$ and $g_2$ of $G$, we usually denote their product
by $g_1g_2$. If we conceive group multiplication from the right as a group action of $G$ on itself, we
will use the notation $g_1 \cdot g_2$ for $g_1g_2$. 
We write $H \leq G$ when $H$ is a subgroup of $G$, and $H < G$ when it is a proper
subgroup. Normal subgroups and proper normal subgroups will be denoted respectively by
$H \unlhd G$ and $H \lhd G$.
For a subset $X$ of $G$, let
$\langle X \rangle$ be the subgroup generated by $X$. The {\em normalizer} of $X$ in $G$ is
$N_G(X) = \{g \in G ~:~ gX=Xg\}.$
For an integer $n$, we denote by $\Z_n$ the group of integers modulo $n$,
and for a prime number $p$, we denote by $\Z_p^*$ the multiplicative group of integers 
relatively prime with $p$. 

The commutator $[x,y]$ of elements
$x$ and $y$ is $x^{-1}y^{-1}xy$. For two subgroups
$X$ and $Y$ of $G$, let $[X,Y]$ be $\langle \{ [x,y] ~:~ x \in X ,y \in Y \} \rangle$.
The derived subgroup $G'$ of $G$ is defined as
$[G,G]$, and its center $Z(G)$ as
$\{z \in G ~:~ gz = zg \mbox{{\rm ~ for all ~}} g \in G \}$. 
The {\em lower central series} of $G$ is the series of subgroups
$G = A_1 \unrhd A_2 \unrhd A_3  \dots,$ where $A_{i+1} = [A_i, G]$ for every $i >1$.
The {\em upper central series} of $G$ is the series of subgroups
$\{1\} = Z_0 \unlhd Z_1 \unlhd Z_2 \ldots,$ where $Z_{i+1} = 
\{x \in G ~:~ [x,g] \in Z_i ~\mbox{for all}~ g \in G\}$ for every $i > 0$.
Clearly $A_2 = G'$ and $Z_1 = Z(G)$. The group $G$ is {\em nilpotent} if there is a natural
number $n$ such that $A_{n+1} = \{1\}$. If $n$ is the smallest integer such that $A_{n+1} = \{1\}$
then $G$ is {\em nilpotent of class $n$}. It is a well known fact that $G$ is nilpotent of class $n$
if and only if $Z_n = G$ in the upper central series. Nilpotent groups of class 1 are simply the nontrivial
abelian groups. A nilpotent group of class at most $n$ is called a {\em nil-$n$} group.

A detailed treatment of nilpotent groups
can be found for example in Hall~\cite{Hall}. Let us just recall here that nilpotent groups are solvable,
and that every $p$-group is nilpotent, where a $p$-group is a finite group whose 
order is a power of some prime number $p$.

\subsection{Nil-$2$ $p$-groups of exponent $p$ }
It is clear from the definition of nilpotent groups that 
$G$ is a nil-2 group if $G'  \leq Z(G).$ It is easy to see that this property implies that
the commutator is a bilinear function in the following sense.

\begin{fact}\label{fact:commutation}
Let $G$ be a nil-$2$ group. Then for every $g_1, g_2, g_3, g_4$ in $G$,
$$ [g_1g_2, g_3g_4] = [g_1,g_2][g_1,g_3][g_2,g_3][g_2,g_4].$$
\end{fact}

The quantum part of our algorithm will deal only with special nilpotent groups
of class 2, which are
also $p$-groups and are of exponent $p$. The structure of these special groups is well known,
and is expressed in the following simple fact.

\begin{fact}\label{fact:nil-2}
Let $G$ be a p-group of exponent $p$ and of nilpotency class $2$. Then
there exist positive integers $m$ and $d$, group elements
$x_1, \ldots , x_m \in G$ and 
$z_1, \ldots, z_d \in G'$ such that 
\begin{enumerate}
\item
${G/G' \cong  \Z_p^m}$ and ${G' \cong \Z_p^d},$
\item
for every $g \in G$ there exists a unique $(e_1, \ldots , e_m, f_1, \ldots f_d) \in \Z_p^{m+d}$ such that
$$ g = x_1^{e_1} \ldots x_m^{e_m} z_1^{f_1} \ldots z_d^{f_d},$$
\item
$G = \langle x_1, \ldots , x_m  \rangle$ and
$G' = \langle z_1, \ldots , z_d  \rangle.$
\end{enumerate}
\end{fact}

We will say that a nil-$2$ $p$-group $G$ of exponent $p$ has parameters $(m,d)$ if 
${G/G' \cong  \Z_p^m}$ and ${G' \cong \Z_p^d}.$ In those groups
we will indentify $G'$ and $\Z_p^d.$ Thus, for two elements $z$ and $z'$ of $G'$, the
product $zz'$ is just $ z \oplus z'$ where $\oplus$ denotes the coordinate-wise addition modulo $p$.
If  $G$ is a such a group
then $|G| = p^{m+d}.$ 
The elements of $G$ can be encoded by binary strings of length
$O((m+d) \log p)$,
and an efficient algorithm on input $G$
has to be polynomial in $m,d$ and $\log p$.

For $j=1, \ldots, p-1$, we consider on generators the maps 
$x_i$ to $x_i^j$. It turns out that these maps extend to automorphisms $\phi_j$ of $G$.
We also define 
the map $\phi_0$ by letting $\phi_0(g)  = 1$, for every $g \in G$.
\begin{proposition}\label{proposition:automorphism}
Let $G$ be a p-group of exponent $p$ and of nilpotency class $2$. Then the mappings
$\phi_j$ have the following properties:
\begin{enumerate}
\item
$\forall j \in \Z_p, \forall z \in G', ~~~ \phi_j(z) = z^{j^2},$
\item
$\forall g \in G, \exists z_g \in G',  \forall j \in \Z_p, ~~ \phi_j(g) = g^jz_g^{j-j^2}.$
\end{enumerate}
\end{proposition}
\begin{proof}
The first statement is trivial when $j=0$. Otherwise,
observe that for every $ j \in \Z_p^*,$ and for every $g\in G$, there exists $z \in G'$ such that
$\phi_j(g) = g^jz$ since $G/G'$ is abelian. To prove the first statement, let $z = [g_1,g_2]$.
Then by this remark, there exist $z_1$ and $z_2$ in $G'$ such that 
$\phi_j([g_1,g_2]) = [g_1^jz_1, g_2^jz_2]$.
By repeated applications of 
Fact~\ref{fact:commutation} this is easily seen to be $([g_1,g_2])^{j^2}.$

We now turn to the second statement.
Let ${j_0}$ be a fixed primitive element of ${\Z_p^*}$. 
Then ${\phi_{j_0}(g)=g^{j_0}  s}$, for some ${s \in G'}$.
Set ${ z_g=s^{(j_0-j_0^2)^{-1}}}$, we have ${\phi_{j_0} (g) = g^{j_0} z_g^{j_0 - j_0^2} }$.
Let ${ k=g  z_{g}}$, then
${\phi_{j_0}(k)= g^{j_0} z_g^{j_0 - j_0^2} z_g^{ j_0^2} 
=k^{j_0}}.$
Therefore, for all $j \in \Z_p$, we have ${ \phi_{j}(k) = k^j}$ and
${ \phi_j(g)=\phi_j(k) \phi_j(z_g^{-1})= g^j z_g^{j}
z_g^{-j^2}}. $

\end{proof}

Clearly, for every $g \in G$, the element $z_g$ whose existence is stated in the second part of 
Proposition~\ref{proposition:automorphism} is unique. From now on, 
let $z_g$ denote this unique element.

\section{Classical reductions in groups of constant nilpotency class}\label{section:redclas}

In order to present the reduction methods
in a sufficiently general way, in this section 
we assume that our groups are presented in terms 
of so-called {\em refined polycyclic presentations} 
\cite{HEO}. Such a presentation of a finite solvable 
group $G$ is based 
on a sequence $G=G_1 \rhd \ldots \rhd G_{s+1}= \{ 1 \},$ where for
each $1\leq i\leq s$ the subgroup $G_{i+1}$
is a normal subgroup of $G_i$ and
the factor group $G_i/G_{i+1}$ is
cyclic of prime order $r_i$. 
For each $i\leq s$ an element 
$g_i\in G_i\setminus G_{i+1}$ is chosen.
Then $g_i^{r_i}\in G_{i+1}$.
Every  element $g$ of $G$ can be uniquely
represented as a product of the
form $g_1^{e_1}\cdots g_s^{e_s}$,
called the normal word for $g$, where
$0\leq e_i<r_i$. 

In the abstract presentation
the generators are $g_1,\ldots,g_s,$
and for each index $1\leq i\leq s,$
the following relations are included:

\begin{itemize}
\item $g_i^{r_i}=u_i$, where 
$u_i=g_{i+1}^{a_{i,i+1}}\cdots g_s^{a_{i,s}}$ 
is the normal word for $g^{r_i}\in G_{i+1},$
\item $g_i^{-1}g_j g_i=w_{ij}$ for every $j>i$, 
where $w_{ij}=g_{i+1}^{b_{i,j,i+1}}\cdots g_s^{b_{i,j,s}}$
is the normal word for $g_i^{-1}g_j g_i\in G_{i+1}$.
\end{itemize}
Using a quantum implementation \cite{ims03}
of an algorithm of Beals and Babai \cite{BB},
refined polycyclic presentation for a solvable black
box group can be computed in polynomial time.
We assume that elements of $G$ are encoded 
by normal words and there is a polynomial time
algorithm in $\log{\size{G}}$, the so called {\em collection procedure},
which computes normal words representing products. 
This is the case for nilpotent groups of constant
class~\cite{Hoefling}. If there is an
efficient collection procedure then refined
polycyclic presentations for subgroups
and factor groups can be obtained
in polynomial time~\cite{HEO}. 
Also, the major notable 
subgroups including Sylow subgroups, 
the center and the commutator can be 
computed efficiently. Furthermore,
in $p$-groups with refined polycyclic presentation,
normalizers of subgroups can be computed in 
polynomial time using the technique of \cite{Eick},
combined with the subspace stabilizer
algorithm of \cite{Luks}.

Our first theorem is a classical reduction for the HSP in groups of constant nilpotency class.
The proof is given by the subsequent three lemmas.

\begin{theorem}\label{theorem:redclas}
Let $\cal C$ be a class of
groups of constant nilpotency class  that is
closed under taking subgroups and 
factor groups. Then the hidden subgroup problem
in members of $\cal C$ can be reduced to the
case where the group is a $p$-group of exponent $p$, and the the subgroup is
either trivial or of cardinality $p$.
\end{theorem}

\begin{corollary}\label{corollary:redclas}
The hidden subgroup problem in
nil-$2$ groups can be reduced to the
case where the group is a $p$-group of exponent $p$, and the the subgroup is
either trivial or of cardinality $p$.
\end{corollary}

\begin{lemma}\label{lemma:redclas1}
Let $\cal C$ be a class of
groups of constant nilpotency class that is
closed under taking subgroups and 
factor groups.
Then the HSP in $\cal C$ can be reduced to the
HSP of $p$-groups belonging to $\cal C$.
\end{lemma}
\begin{proof}
As a nilpotent group $G$ is the direct product
of its Sylow subgroups, any subgroup
$H$ of $G$ is the product of its intersections
with the Sylow subgroups of $G$. 
\end{proof}

\begin{lemma}\label{lemma:redclas2}
Let $\cal C$ be a class of
$p$-groups of constant nilpotency class that is
closed under taking subgroups and 
factor groups. Then the hidden subgroup problem
in members of $\cal C$ can be reduced to the
case where the subgroup is
either trivial or of cardinality $p$.
\end{lemma}
\begin{proof}
Assume that we have a procedure $\cal P$ which finds
hidden subgroups in  $\cal C$ under the
promise that the hidden subgroup is trivial or is of order $p$. 
Let $G$ be a group in $\cal C$ and let $f$ be 
a function on $G$ hiding the subgroup $H$ of $G$.
We describe an iterative procedure which uses
$\cal P$ as a subroutine and finds $H$ in $G$. The basic idea is to
compute a refined polycyclic sequence $G = G_1 \rhd \ldots \rhd G_s \rhd 1$
for $G$ and to proceed  calling $\cal P$ on the subgroups in the sequence starting with $G_s$.
When $\cal P$ finds for the first time a nontrivial 
subgroup generated by $h$, then we would like to restart
the process in $G/ \langle h\rangle $, and at the end, collect all the generators.
Since $\langle h \rangle $ is not necessarily a 
normal subgroup of $G$ we will actually restart the process
instead in $N_G(\langle h\rangle)$.

More formally, let us suppose that $f$ hides $H$ in $G$, and let $\widetilde H$ be some
subgroup of $H$. Then $f$ hides $N_G(\widetilde H)\cap H$ in 
$N_G(\widetilde H)$, and therefore it hides $(N_G(\widetilde H)\cap H)/\widetilde H$
in $N_G(\widetilde H) /\widetilde H$.
We consider the following algorithm: \\


\begin{algorithm}
\caption{}
\label{algo:divideconquer}
\begin{algorithmic}

\STATE success:= TRUE, $\widetilde H=\{1\}$.
\WHILE{success=TRUE} 
 \IF{$G \neq \widetilde H$} 
  \STATE compute $N_G(\widetilde H)/\widetilde H = G_1 \rhd \ldots \rhd G_s \rhd 1$ a
       refined polycyclic representation, $i:=s$
          \WHILE{$ i > 0$}
             \STATE call $\cal P$  on $G_i$
             \IF{$\cal P$ returns $\langle h \rangle $}
                \STATE $\widetilde H := \langle \widetilde H \cup \{h\} \rangle, i=0 $
             \ELSE 
                 \STATE $i:=i-1$
                 \IF{$i=0$}  \STATE success := FALSE
                 \ENDIF
             \ENDIF
           \ENDWHILE
  \ELSE \STATE success:=FALSE
  \ENDIF
\ENDWHILE

\end{algorithmic}
\label{algo:minimum}
\end{algorithm}

\suppress{
success:= TRUE, $\widetilde H=\{1\}$.\\
while success=TRUE do\\
\ compute $N_G(\widetilde H)/\widetilde H = T_1>\ldots>T_s>1$ a
refined polycyclic representation\\
$\ \ \ i:=s$\\
while $ i > 0$ do \\
call $\cal P$  on $T_i$\\
if $\cal P$ returns $<h>$ then $\widetilde H := < \widetilde H \cup \{h\}>, i=0 $\\
else $i:=i-1$\\
if $i=0$ then success = FALSE\\
}

Algorithm 1 stops when the subgroup $\widetilde H$ is such that 
$(N_G(\widetilde H) \cap H )/\widetilde H   = \{1\}$, that is when 
$N_G(\widetilde H) \cap H = \widetilde H$. We claim that this implies
$\widetilde H = H$. Indeed, suppose that $\widetilde H$ is a proper
subgroup of $H$. Since in nilpotent groups a proper subgroup is also a proper
subgroup of its normalizer, $\widetilde H$ is also a proper subgroup of 
$N_H(\widetilde H) = N_G(\widetilde H) \cap H $.
 
Finally observe that the whole process makes 
$O(\log_p^2{\size{G}})$ calls to $\cal P$.

\end{proof}

\begin{lemma}\label{lemma:redclas3}
Let $\cal C$ be a class of
$p$-groups of constant nilpotency class that is
closed under taking subgroups and 
factor groups. Then the instances of the hidden subgroup problem
in members of $\cal C$, when the subgroup is either trivial or of cardinality $p,$
can be reduced to 
groups in $\cal C$ of exponent $p$.
\end{lemma}

\begin{proof}

If $p$ is not larger than
the class of $G$, the algorithm of 
\cite{fimss03} is applicable.
Otherwise
the elements of order $p$ or 1
form a subgroup $G^*$, see~Chapter 12
of~\cite{Hall}. The hidden subgroup $H$ is also a subgroup of $G^*$ since $|H| \leq p$.
The function hiding $H$ in $G$ also hides it in $G^*$, therefore the reduction will
consist in determining $G^*$.

We design an algorithm that finds $G^*$ 
by induction on the length of refined polycyclic presentations.
If $|G| = p$ then $G^* = G$.  Otherwise, let 
$G=G_1 \rhd G_2 \rhd  \ldots \rhd G_{s} \rhd \{1\}$ be
a refined polycyclic presentation with $s \geq2$.
It is easy to construct a presentation where $G_s$ is a subgroup of the center of $G$, 
which we suppose from now on. For the ease of notation we set  $M= G_2$ and $N=G_s$.

We first describe the inductive step in a simplified case, with the additional hypothesis $(G/N)^* = G/N$.
Observe that the hypothesis is equivalent to saying that the map $\phi:x\mapsto x^p$
sends every element of $G$ into $N$. From this it is also clear that the hypothesis carries
over to $M$, that is $(M/N)^* = M/N$. We further claim that
either $G^*=G$ or $G^*$ is a subgroup of $G$
of index $p$.
In fact this follows Theorem  12.4.4 of~\cite{Hall} which states that the map $\phi$ is constant
on cosets of $G^*$ and distinct on different cosets.
>From a polycyclic presentation
of $G$ it can be read off whether or not $G=G^*$. If $G^* = G$ we are done.
Otherwise we compute inductively $M^*$.
If $M^* = M$ then $G^* = M$. If $M^*$ is a proper subgroup of $M$
then $M^*$  has index $p^2$ in $G$. 
Pick an arbitrary $u\in M\setminus M^*$ and $y \in G \setminus M$.
By the assumptions, $u^p=g_s^{j_u}$ for some integer $0<j_u<p,$
and $y^p=g_s^{j_y}$ for some integer $0\leq j_y<p$. Recall that in 
the polycyclic presentation model, computing
normal words for $u^p$ and $y^p$ -- using fast exponentiation --
amounts to computing $j_u$ and $j_y$. 
Set $x=u^{j_y j_u^{-1}}$. For this $x$ we have 
$x^p = y^p$, and therefore
$xy^{-1} \in G^*$.
Since $xy^{-1} \in G^* \setminus M^*$, we have
$G^* = \langle M^*, xy^{-1} \rangle $.

In the general case first $(G/N)^*$ is computed inductively. If $(G/N)^* = G/N$ then one proceeds
as in the simplified case.
Otherwise we set
$K = (G/N)^* N$. We claim that $G^* = K^*$. For this we will show that.
 $G^* \subseteq K$. To see this, let $x$ be an element of $G^*$.
Then $x = yz$ where $y \in G/N$ and $z \in N$. We show that $y$ is in $(G/N)^*$ which implies
that $x \in K$. Indeed, $y^p = y^pz^p=(yz)^p=1$, where the first equality follows from $|N| =p$, the 
second from $N \leq Z(G)$ and the third from $x \in G^*$. Finally observe that $(K/N)^* = K/N$ since
$K/N = (G/N)^*$. Therefore one can determine $K^*$ inductively as in the simplified case.

Let $c(s)$ denote the number of recursive calls when the length of a presentation is $s$.
In the simplified case the number of calls is $s-1$. Therefore in the general case we have
$c(s) = c(s-1) + s-2$, whose solution is $c(s) = O(s^2)$.

\end{proof}

\section{The quantum algorithm}\label{section:algorithm}
The quantum part of our algorithm, up to technicalities, follows the same lines as 
the algorithm given in~\cite{iss07}
for extraspecial groups. %
The proof in these section are included here for the sake of completeness.

\begin{theorem}\label{theorem:threelemmas}
Let $G$ be a nil-$2$ $p$-group of exponent $p$, and 
let us given an oracle $f$ which
hides a subgroup $H$ of $G$ whose cardinality is either 1 or $p$.
If we have an efficient quantum procedure (using $f$)
which hides $HG'$ in $G$ then $H$ can be found efficiently.
\end{theorem}

\begin{proof}
First observe that finding $H$ is efficiently reducible to finding $HG'$.
Indeed, $HG'$ is an abelian subgroup of $G$ since $H$ is abelian.
The restriction of the hiding function $f$ to  $HG'$
of $G$ hides $H$.
Therefore the standard algorithm for solving the $\HSP$ in abelian groups
applied to $HG'$ with oracle $f$ yields $H$.

Let us now suppose that $G$ has parameters $(m,d).$
We will show that finding $HG'$ can be efficiently reduced to 
the hidden subgroup problem in an abelian group.
Let us denote for every element  
$ g = x_1^{e_1} \ldots x_m^{e_m} z_1^{f_1} \ldots z_d^{f_d}$ of $G$, 
by $\overline{g}$
the element $x_1^{e_1} \ldots x_m^{e_m}$.
We define the group $\overline{G}$ whose base set is $\{ \overline{g} : g \in G\}$. Observe that
this set of elements does not form a subgroup in $G$.
To make $\overline{G}$ a group, its law is defined by
$\overline{g_1} \ast \overline{g_2}=\overline{g_1  g_2}$
for all $\overline{g_1}$ and $\overline{g_2}$ in $\overline{G}$.
It is easy to check that $\ast$ is well defined,
and is indeed a group multiplication. In fact, the group $\overline{G}$ is
isomorphic to $G/G'$ and therefore is isomorphic to $\Z_p^{m}$.
For our purposes a nice way to think about
$\overline{G}$ as a representation of $G/G'$ with unique encoding. Observe also that 
$HG' \cap \overline{G}$ is a subgroup of $(\overline{G}, \ast)$ because
$HG'/G'$  is a subgroup of $G/G'$.
Since $HG'=(HG' \cap \overline{G}) G'$, 
finding $HG'$ is efficiently reducible to finding $HG' \cap \overline{G}$ in $\overline{G}$.

To finish the proof, let us remark that
the procedure which hides $HG'$ in $G$ hides also
$HG' \cap \overline{G}$ in $\overline{G}$. Since $\overline{G}$ is abelian,
Fact~\ref{fact:hsp} implies that we can find efficiently
$HG{'} \cap \overline{G}$.
\end{proof}

\begin{theorem}\label{theorem:HG'}
Let $G$ be a nil-$2$ $p$-group of exponent $p$, and 
let us given an oracle $f$ which
hides a subgroup $H$ of $G$.
Then there is an efficient quantum procedure which
hides $HG'$ in $G$.
\end{theorem}
\begin{proof}

The basic idea of the quantum procedure is the following. 
Suppose that we could create, for some $a \in G$,
the coset state $\ket{aHG'}$. Then the group action $g \rightarrow \ket{aHG' \cdot g}$
is a hiding procedure. Unfortunately, $\ket{aHG'}$ can only be created efficiently when 
$p$ and $d$ are constant.
In general, we can create efficiently $\ket{aHG'_u}$ for random $a \in G$ and $u \in G'$, where
by definition $\ket{G'_u}=
\frac{1}{\sqrt{|G'|}} \sum_{z \in \Z_p^d} \omega^{-<u,z>} \ket{z}$. Then
$\ket{aHG'_u \cdot h} = \ket{aHG'_u}$ for every $h \in H$, and
$\ket{G'_u \cdot  z}=\omega^{<u,z>} \ket{G'_u}$.
To cancel the disturbing phase we will use more sophisticated group action via the group
automorphisms $\phi_j$ on several copies of the states $\ket{aHG'_u}$.

\begin{lemma}\label{lemma:superposition}
There is an efficient quantum procedure which creates
$ \frac{1}{\sqrt{p^d}} \sum _{u \in \Z_p^d} \ket{u} \ket{aHG'_u}$ where $a$
is a random element from $G$.
\end{lemma}
\begin{proof}
We start with $\ket{0}\ket{0}\ket{0}$.
Since we have access to the hiding function $f$,
we can create the superposition $\frac{1}{\sqrt{|G|}}\sum_{g \in G} \ket{0} \ket{g} \ket{f(g)}$.
Observing and discharging the third
register we get $\ket{0} \ket{aH}$ for a random element~$a$.
Applying the Fourier transform over $\Z_p^d$ to the
first register gives $\ket{\Z_p}\ket{aH}$.
Multiplying  the second register by the opposite of the first one results in
$\frac{1}{\sqrt{p^d}} \sum _{z \in \Z_p^d} \ket{-z} \ket{a H z}$. A final Fourier transform in the first register
creates the required superposition.
\end{proof}

Our next lemma which is an immediate consequence of Proposition~\ref{proposition:automorphism}
claims that the states $\ket{aHG'_u}$ are eigenvectors of
the group action of multiplication from the right by $\phi_j(g)$,
whenever $g$ is from $HG'$. Moreover, the
corresponding eigenvalues are some powers of the
root of the unity, the exponent does not depend on $a$, and the dependence on $u$ and $j$
is relatively simple.

\begin{lemma}\label{lemma:action}
We have
\begin{enumerate}
\item $\forall z \in \Z_p^d, \forall a \in G, \forall u \in \Z_p^d, \forall j \in \Z_p,~~
\ket{aHG'_u \cdot \phi_j(z)}=\omega^{<u,z> j^2} \ket{aHG'_u},$
\item  $\forall h \in H,  \forall a \in G, \forall u \in \Z_p^d, \forall j \in \Z_p, ~~
\ket{aHG'_u \cdot  \phi_j(h)} ~=~ \omega^{<u,z_h> (j-j^2)} \ket{aHG'_u}.$
\end{enumerate}
\end{lemma}

The principal idea now is to
take several copies of the states $\ket{a_iHG'_{u_i}}$ and choose the
$j_i$ so that the product
of the corresponding
eigenvalues
becomes
the unity. Therefore the combined actions
$\phi_{j_i}(g)$, when $g$ is from $HG'$, will not modify the combined state. It turns out
that we can achieve this with a sufficiently big enough number of copies. Let $n = n(d)$ 
some function of $d$
to be determined later.

For $ \overline{a} = (a_1, \ldots, a_n) \in G^n, 
~ \overline{u} = (u_1, \ldots , u_n) \in ({\Z_p^d})^n,
~ \overline{j} = (j_1, \ldots, j_n) \in ({\Z_p})^n \setminus \{{0^n}\}$ 
and $g \in G$, we define the quantum state
$\ket{ \Psi_g^{\overline{a}, \overline{u}, \overline{j}}}$ in $\C^{G^n}$ by
$$ \ket{ \Psi_g^{\overline{a}, \overline{u}, \overline{j}}}  =
 \bigotimes_{i=1}^n    \ket{a_iHG'_{u_i} \cdot \phi_{j_i}(g)} .$$

Our purpose is to find an efficient procedure to generate triples
$(\overline{a}, \overline{u}, \overline{j})$ such that
for
every $g$ in $HG'$, $ \ket{ \Psi_g^{\overline{a}, \overline{u}, \overline{j}}}  =
\bigotimes_{i=1}^n    \ket{a_iHG'_{u_i}}  $. We call such triples
{\em appropriate}. The reason to look for appropriate triples is that they lead to hiding
sets for $HG'$ in $G$ as stated in the next lemma.

\begin{lemma}\label{lemma:appropriate}
If $(\overline{a}, \overline{u}, \overline{j})$ is an appropriate triple then
$\{ \ket{ \Psi_g^{\overline{a}, \overline{u}, \overline{j}}} : g \in G \}$ is hiding for $HG'$ in $G$.
\end{lemma}

\begin{proof}
To see this, first observe that $HG'$ is a normal subgroup of $G$.
If $g_1$ and $g_2$ are in different cosets of $HG'$ in $G$ then
let $1 \leq i \leq n$ such that  $j_i \neq 0$.
The elements $\phi_{j_i}(g_1)$ and $\phi_{j_i}(g_2)$ are in different cosets of
$HG'$ in $G$ since $\phi_{j_i}$ is an automorphism of $G$. Also, we have
$\supp(\ket{aHG'_u}) =  \supp (  \ket{aHG'})$, and therefore
$\supp(\ket{aHG'_u \cdot \phi_{j_1}(g_1)})$ and
$\supp(\ket{aHG'_u \cdot \phi_{j_2}(g_2)})$ are included in different cosets and are disjoint.
Thus
the states $\ket{ \Psi_{g_1}^{\overline{a}, \overline{u}, \overline{j}}}$ and
$\ket{ \Psi_{g_2}^{\overline{a}, \overline{u}, \overline{j}}}$ are orthogonal.

If $g_1$ and $g_2$ are in the same coset of $HG'$ then $g_1 = g g_2$ for some $g \in HG'$,
and for all $1 \leq i \leq n,$ we have $\phi_{j_i}(g_1) = \phi_{j_i}(g) \phi_{j_i}(g_2)$. Thus
$\ket{ \Psi_{g_1}^{\overline{a}, \overline{u}, \overline{j}}} =
\ket{ \Psi_{g g_2}^{\overline{a}, \overline{u}, \overline{j}}}  =
\ket{ \Psi_{g_2}^{\overline{a}, \overline{u}, \overline{j}}}$.
\end{proof}

Let us now address the question of existence of appropriate triples and efficient ways to
generate them. Let $(\overline{a}, \overline{u}, \overline{j})$ be an arbitrary element of
$G^n \times (\Z_p^d)^n \times (\Z_p)^n \setminus \{{0^n}\}$, and
let $g$ be an element of  $HG'$. Then $g=h z$ for some $h \in H$ and 
$z \in \Z_p^d$, and $\phi_{j_i}(g)=\phi_{j_i}(h) \phi_{j_i}(z)$ for $i=1, \ldots, n$.
By Lemma~\ref{lemma:action}, we have
$\ket{a_iHG'_{u_i} \cdot \phi_{j_i}(z)}= \omega^{<u_i,z> j_i^2 }\ket{a_iHG'_{u_i}}$, and
$\ket{a_iHG'_{u_i} \cdot \phi_{j_i}(h)}= 
\omega^{<u_i,z_h> (j_i-j_i^2)}\ket{a_iHG'_{u_i}}$, and
therefore
$$
 \ket{ \Psi_g^{\overline{a}, \overline{u}, \overline{j}}}
= \omega^{\sum_{i=1}^n  <u_i,z_h> (j_i-j_i^2) +  <u_i,z> j_i^2  }
\bigotimes_{i=1}^n    \ket{a_iHG'_{u_i}}.
$$
For a given $\overline{u}$, we consider the following system of
quadratic equations, written in vectorial form:
\begin{equation*}
\begin{cases}
\sum_{i=1}^n  u_i(j_i-j_i^2) & =~~ 0^d \\
\sum_{i=1}^n  u_ij_i^2  & =~~ 0^d.
\end{cases}
\end{equation*}
It should be clear that when this system has a nontrivial solution $\overline{j}$
(that is $ \overline{j} \neq 0^d$) then
$(\overline{a}, \overline{u}, \overline{j})$ is an appropriate triple, for every $\overline{a}$.
In fact, the Chevalley-Warning theorem~\cite{c36, w36} implies that 
the following equivalent system of vectorial equations
has a nontrivial solution for every $\overline{u}$, whenever $n > 3d$.
\begin{equation}\label{original}
\begin{cases}
\sum_{i=1}^n  u_ij_i^2 & =~~ 0^d \\
\sum_{i=1}^n  u_ij_i  & =~~ 0^d.
\end{cases}
\end{equation}
Moreover, if we take a substantially larger number of variables, we can find a solution
in polynomial time.
\begin{theorem}\label{theorem:equation}
If $n = (d+1)^2(d+2)/2$ then we can find a nontrivial solution for the system~(\ref{original})
in polynomial time.
\end{theorem}
The proof of Theorem~\ref{theorem:equation} will be given
in the next section.
To finish the proof of Theorem~\ref{theorem:HG'} we describe the efficient hiding procedure.
On input $\ket{g}$, 
it computes, for some $\overline{a} \in G^n$, the superposition
$$ \frac{1}{{p}^d} \bigotimes_{i=1}^n   \sum _{u_i \in \Z_p} \ket{u_i} \ket{a_iHG'_{u_i}},$$
which by Lemma~\ref{lemma:superposition} can be done efficiently, and then
it measures the registers for the $u_i$. Then, by Theorem~\ref{theorem:equation} it
finds efficiently a nontrivial solution $ \overline{j}  $ for system~(\ref{original}).
Such a triple $(\overline{a}, \overline{u}, \overline{j})$ is appropriate, and therefore by
Lemma~\ref{lemma:appropriate}
$\{ \ket{ \Psi_g^{\overline{a}, \overline{u}, \overline{j}}} : g \in G \}$ is hiding for $HG'$ in $G$.
Using the
additional input $\ket{g}$, the procedure finally computes
$ \ket{ \Psi_g^{\overline{a}, \overline{u}, \overline{j}}} $.
\end{proof}

\section{Solving the system of equations}\label{section:equation}
This section is fully dedicated to the proof of Theorem~\ref{theorem:equation}.
If $p= 2$ then the $d$ quadratic and the $d$ linear equations 
coincide, and the (linear) system can easily be solved in polynomial time. Therefore,
from now on, we suppose that $p > 2$.
Let us detail system~(\ref{original}), where we set 
$u_i = (u_{1,i}, u_{2,i}, \ldots , u_{d,i} ).$ We have the
following system of $d$ homogenous quadratic  and $d$ homogenous linear one
equations with $n$ variables:

\begin{equation}\label{detailed}
\begin{cases}
\forall \ell \in [ | 1, d | ],~~~\sum_{i=1}^n  u_{\ell,i}j_i^2 & =~~ 0 \\
\forall \ell \in [ | 1, d | ],~~~\sum_{i=1}^n  u_{\ell,i}j_i  & =~~ 0\\
\end{cases}
\end{equation}

We start by considering only the quadratic part of the~(\ref{detailed}), that is 
\begin{equation}\label{quadratic}
\begin{cases}
\forall \ell \in [ | 1, d | ],~~~\sum_{i=1}^{n'}  u_{\ell,i}j_i^2 & =~~ 0 \\
\end{cases}
\end{equation} 
for some integer $n'$.

\begin{claim}\label{claim:quadratic}
If $n' = (d+1)(d+2)/2$ then we can find a nontrivial solution for~(\ref{quadratic}) in polynomial time.
\end{claim}

\begin{proof}
For the ease of notation we 
are going to represent this system by the $d \times n'$ matrix

$$M = \begin{pmatrix}
u_{1,1} & \ldots & u_{1,n'} \\
\vdots  &        & \vdots \\
u_{d,1} & \ldots & u_{d,n'}
\end{pmatrix}.$$
We will present a recursive algorithm whose complexity will be polynomial in $d$ and in $\log p$.
When $d=1$, the unique quadratic equation is of the form 
$u_{1,1} j_1^2 + u_{1,2} j_2^2+ u_{1,3} j_3^2=0$. According to a special case
of the main result in the thesis of 
van de Woestijne  (Theorem A3 of~\cite{w06}), a nontrivial solution for this can be 
found in polynomial time in $\log p$.

Let us suppose now that we have $d$ equations in $n' = (d+1)(d+2)/2$ variables.
We can make elementary operations on $M$ (adding two lines and multiplying a line with a nonzero 
constant) without changing the solutions of the system.
Our purpose is to reduce it with such operations 
to $d-1$ equations in
at least $d(d+1)/2$ variables.
If the system is of rank less than $d$, then we can erase an equation and get an 
equivalent system with only $d-1$ equations in the same number of variables. Otherwise,
we perform Gaussian elimination resulting in the matrix

$$M_1 = \begin{pmatrix}
1 & 0 & 0 & \ldots & 0 & u_{1,d+1}^{(1)} & \ldots & u_{1,n'}^{(1)} \\
0 & 1 & 0 & \ldots & 0 & u_{2,d+1}^{(1)} & \ldots & u_{2,n'}^{(1)} \\
\vdots  &  & \ddots &   & \vdots &  \vdots & & \vdots \\
0 & \ldots & 0 & 1 & 0 & u_{d-1,d+1}^{(1)} & \ldots & u_{d-1,n'}^{(1)}\\
0 & \ldots & 0 & 0 & 1 & u_{d,d+1}^{(1)} & \ldots & u_{d,n'}^{(1)}
\end{pmatrix}.$$

Since checking quadratic residuosity is simple, and for
odd $p$, half of the elements of $\Z_p^*$ are squares, we can easily compute a quadratic non-residue
$\lambda$ in probabilistic polynomial time.
Then every quadratic non-residue is the product of a square and $\lambda$.
We will look at column $d+1$ of $M_1.$
If the column is everywhere 0 then $j_{d+1} = 1$ and $j_i = 0$ for $i \neq d+1$ is a nontrivial
solution of the whole system. Otherwise, 
without loss of generality, we can suppose that
for some $(k_1,k_2) \neq (0,0)$ the first $k_1$ elements are squares, 
the following $k_2$ elements are the product of $\lambda$ and a square, 
and the last $d-k_1-k_2$ elements are zero. Thus there exist $v_1, \ldots , v_{k_1+k_2}$
different from 0, 
such that $u_{i,d+1}^{(1)} = v_i^2$ for $1 \leq i \leq k_1,$ and 
$u_{i,d+1}^{(1)} = \lambda v_i^2$ for $k_1 + 1 \leq i \leq k_1 + k_2.$
Once we have a quadratic non-residue, 
the square roots $v_1, \ldots , v_{k_1+k_2}$ can be found in deterministic polynomial time
in $\log p$ by the Shanks-Tonelli algorithm~\cite{st72}.
We set the variables $j_{k_1+k_2+1}, \ldots , j_d$ to 0, and eliminate columns
${k_1+k_2+1}, \ldots , d$ from $M_1$. Then for $i = 1, \ldots , k_1 + k_2,$ we divide the line $i$
by $v_{i}^2$. Introducing the new variables $j_i' = j_i v_i^{-1}$ for $ 1 \leq i \leq k_1 + k_2$,
the matrix of the system in the $n' - d + k_1 + k_2$ variables 
$j_1', \ldots, j_{k_1+k_2}' , j_{d+1}, \ldots j_{n'}$ is

$$M_2 = \left(\begin{array}{cccccccccc}
1 & 0&  & \ldots & & 0& 1& u_{1,d+2}^{(2)} & \ldots & u_{1,n'}^{(2)} \\
0&\ddots & &&& & \vdots &  \vdots & & \vdots \\
&&1&\ddots&&\vdots& 1& u_{{k_1},d+2}^{(2)} & \ldots & u_{{k_1},n'}^{(2)} \\
\vdots&&\ddots&1&&& \lambda & u_{{k_1+1},d+2}^{(2)} & \ldots & u_{{k_1+1},n'}^{(2)} \\
&&&&\ddots& 0& \vdots &  \vdots & & \vdots \\
0&&\ldots&&0&1& \lambda & u_{{k_1 + k_2},d+2}^{(2)} & \ldots & u_{{k_1 + k_2},n'}^{(2)} \\
0&&&\ldots&&& 0& u_{{k_1 + k_2 + 1 },d+2}^{(2)} & \ldots & u_{{k_1 + k_2 + 1},n'}^{(2)} \\
\vdots&&&&&\vdots& \vdots &  \vdots & & \vdots \\
0&&&\ldots&&&0& u_{d,d+2}^{(2)} & \ldots & u_{d,n'}^{(2)}
\end{array}\right).$$

In $M_2$ we subtract the first line from lines $2, \ldots, k$ and line 
$k_1 + 1$ from lines $k_1 +2, \ldots , k_1 + k_2$. Then we set the variables 
$j_2', \ldots ,  j_{k_1}'$ to $j_1'$, and variables 
$j_{k_1 + 2}', \ldots ,  j_{k_1 + k_2}'$ to $j_{k_1 + 1}'$. The corresponding
changes in the matrix are eliminating columns $2, \ldots  {k_1}$
and ${k_1 + 2}, \ldots  {k_1 + k_2}$ and putting in  columns 1 and $k_1 + 1$  everywhere 0
but respectively in line 1 and line $k_1 + 1$.
Finally, by exchanging line $2$ and line $k_1+1$, we get the matrix

$$M_3 = \begin{pmatrix}
1 & 0 & 1 & u_{1,d+2}^{(3)} & \ldots & u_{1,n'}^{(3)} \\
0 & 1 & \lambda & u_{2,d+2}^{(3)} & \ldots & u_{2,n'}^{(3)} \\
0&  0 & 0 &u_{3,d+2}^{(3)} & \ldots & u_{3,n'}^{(3)} \\
\vdots & \vdots & \vdots &  \vdots & & \vdots \\
0&0 & 0& u_{d,d+2}^{(3)} & \ldots & u_{d,n'}^{(3)}
\end{pmatrix}$$
in variables $j_1', j_{k_1+1}', j_{d+1}, \ldots , j_{n'}$.

To finish the reduction, we will distinguish two cases, depending on the congruency class of $p$
modulo 4.  When $p \equiv 1$, the element $-1$ is a square, and in polynomial time in
$\log p$ we can find $s$ such that $s^2 = -1$. We set $j_1 = s j_{d+1}$, eliminate column 1
from matrix $M_3$, put 0 in line 1 column $d+1$, and exchange line 1 and line 2.
When $p \equiv 3$ modulo $4$, the element $-1$ is not a square, and therefore we can choose
$\lambda = -1$. We set $j_2 = j_{d+1}$, eliminate column 2, and put 0 in line 2 column $d+1$.

In both cases we end up with a matrix of the form

$$M_4 = \begin{pmatrix}
1 & \alpha & u_{1,d+2}^{(3)} & \ldots & u_{1,n'}^{(3)} \\
0 & 0 & u_{2,d+2}^{(3)} & \ldots & u_{2,n'}^{(3)} \\
\vdots & \vdots & \vdots &   &  \vdots \\
0 & 0& u_{d,d+2}^{(3)} & \ldots & u_{d,n'}^{(3)}
\end{pmatrix}$$
in the variables $j',  j_{d+1}, \ldots , j_{n'}$
where $\alpha = \lambda$ and  $j' =  j_{k_1+1}'$
when $p \equiv 1$, and $\alpha = 1$  and 
$j' = j_1'$ otherwise. Without the first line it represents a system 
of $d-1$ equations in $n' - (d+1) = d(d+1)/2$ variables for which we can find
a nontrivial solution by induction.
Let $ j_{d+2}, \ldots , j_{n'}$ such a solution, and set $b = \sum_{k = d+2}^{n'} u_{1,k}^{(3)} j_k$.
To give values to the remaining two variables we have to solve the equation
$j'^2+ \alpha j_{d+1}^2 + b =0$. It is easy to see that the equation is always solvable, and
then by Theorem A3 of~\cite{w06} a solution can be found deterministically in polynomial time.

Gaussian elimination on $M$ can be done in time $O(d^4).$ Finding a nontrivial solution 
for a quadratic homogeneous equation in 3 variables takes time $q_1(\log p)$,
solving a quadratic equation in two variables takes time $q_2(\log p)$,
and finding a square roots modulo $p$ takes time $q_3(\log p)$ where $q_1, q_2$ and $q_3$
are polynomials. Therefore the complexity of solving system~(\ref{original}) is
$O(d^5 + d^2 q_3(\log p) + d q_2(\log p) + q_1(\log p) )$.

\end{proof}

We now turn to the system~(\ref{detailed}).
Let $n' = n/(d+1)$, and for $0 \leq k \leq d,$ consider the the system
of $d$ quadratic equations in $n'$ variables:

\begin{equation*}
\begin{cases}
\forall \ell \in [ | 1, d | ],~~~\sum_{i=kn'+1}^{(k+1)n'}  u_{\ell,i}j_i^2 & =~~ 0. \\
\end{cases}
\end{equation*} 
By Claim~\ref{claim:quadratic}, each of these systems has a nontrivial solution that we can
find in polynomial time. For each $k$,
let $(j_{kn'+1}, \ldots, j_{(k+1)n'})$ such a solution of the $k$th quadratic system.
Then the set
$$ \{(\lambda_0 j_1, \ldots, \lambda_0 j_{n'}, \lambda_1 j_{n'+1}, \ldots, \lambda_1 j_{2n'}, \ldots,
\lambda_d j_{dn'+1}, \ldots, \lambda_d j_{(d+1)n'}) ~:~ 
(\lambda_0, \lambda_1, \ldots, \lambda_d) \in \Z_p^{d+1} \}$$ 
is a $d+1$ dimensional subspace of of $\Z_p^n$ 
whose elements are solutions of the  $d$ quadratic equations
in~(\ref{detailed}). Since in (\ref{detailed}) there are $d$ linear equations, we can find a
a nontrivial $(\lambda_0, \lambda_1, \ldots, \lambda_d) \in \Z_p^{d+1}$ such that
$(\lambda_0 j_1, \ldots, \lambda_0 j_{n'}, \lambda_1 j_{n'+1}, \ldots, \lambda_1 j_{2n'}, \ldots,
\lambda_d j_{dn'+1}, \ldots, \lambda_d j_{(d+1)n'})$ is a (nontrivial) solution of the linear
part of~(\ref{detailed}), and therefore of the whole system. \qed

Observe that the only probabilistic part of the algorithm is the generation of a quadratic non-residue
modulo $p$.


\end{document}